# Thermally-driven flows between a Leidenfrost solid and a ratchet surface

Steffen Hardt[1], Sudarshan Tiwari[2], Tobias Baier[1]

[1]Center of Smart Interfaces, TU Darmstadt, Petersenstr. 32, D-64287 Darmstadt, Germany

[2]Fachbereich Mathematik, TU Kaiserslautern, Erwin-Schrödinger-Str., D-67663 Kaiserslautern, Germany

**Abstract**

The significance of thermally-driven flows for the propulsion of Leidenfrost solids on a ratchet surface is studied based on a numerical solution of the Boltzmann equation. In contrast to a previous analysis, it is found that no significant thermal creep flow is established. Instead, the flow pattern is dominated by thermal edge and thermal-stress slip flow, the latter being directed opposite to thermal creep flow. However, in total thermally-induced flows only make a minor contribution to the propulsion of Leidenfrost solids on ratchet surfaces which is dominated by the pressure-driven flow due to the sublimating solid.

Droplets deposited on a ratchet surface heated to temperatures above the Leidenfrost point [1, 2] exhibit a directed motion, as was first discovered by Linke et al. in 2006 [3]. In corresponding experiments a droplet hovers above a microstructured surface equipped with parallel, asymmetric grooves, resting on a cushion of its own vapor. A droplet motion normal to the orientation of the grooves is observed. Pinpointing the mechanism of droplet propulsion is a quite challenging task, since a variety of different effects come into consideration. The most prominent ones are viscous shear, net pressure forces due to the drop surface following the ratchet's contour, thermocapillary flows, gradients in Laplace pressure, coupling of surface waves and droplet oscillations to the translational motion, recoil pressure due to phase change, and thermal creep flow [3-6]. In that context, a major step forward was the observation that sublimating Leidenfrost solids such as disks of dry ice are propelled on ratchet surfaces in a similar manner as Leidenforst droplets [4]. This excludes many of the effects listed above (such as thermocapillary flows), but still does not allow to clearly pinpoint the propulsion mechanism. In that context there are at least two different pictures that have been suggested to explain the propulsion of Leidenfrost solids. First of all, there are experimental hints that viscous stresses originating from the flow of sublimating vapor guided over the topography of the microstructured surface result in a significant net force on the solid hovering above the surface [5]. Alternatively, it was suggested that thermal creep flow developing inside the narrow gap between the Leidenfrost solid and the surface could be the origin of the propulsion mechanism [6]. These are two completely different scenarios, since the former picture suggests that the production of vapor is the cause of the net force, whereas in the latter a motion could be induced even without any vapor influx into the gap between the surfaces in situations where their separation is not maintained by a cushion of vapor due to sublimation. The purpose of this letter is to examine the nature of thermally-driven flows in that context and to quantify their contribution to the propulsion of Leidenfrost solids. For this reason the flux of sublimated vapor due to sublimation is initially neglected.

The corresponding model geometry is shown in Fig. 1. The gap between a ratchet surface and a planar surface is filled with a gas. The ratchet surface is periodic in $x$-direction and translationally-invariant in $y$-direction. Therefore, it is sufficient to consider a unit cell of extension $L$ in $x$-direction, as indicated by the dashed rectangle. Each of the two surfaces is assumed to be isothermal, the upper surface representing the bottom of the Leidenfrost solid and having a temperature of $T_{subl}$, the lower one a temperature of $T_0$. The model geometry of Fig. 1 closely resembles the surface topographies that have been used to experimentally study the propulsion of Leidenfrost objects on ratchet surfaces.

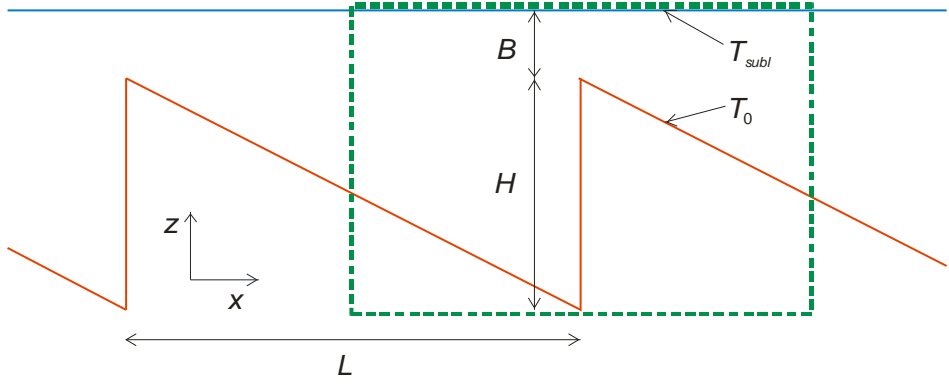

FIG. 1. (Color online) Model geometry considered for studying thermally-driven gas flows between a Leidenfrost solid and a ratchet surface. The blue surface represents the bottom of the Leidenfrost solid, the red one the ratchet surface.

Thermally-driven gas flows are phenomena that cannot be captured from first principles using the Navier-Stokes equations. Therefore, the Boltzmann equation

$$\frac{\partial}{\partial t}f + \mathbf{u}\cdot\nabla_{\mathbf{r}}f = J_{HS}[f] \qquad (1)$$

is employed to model the transport processes occurring inside the gas, where it is assumed that no external forces are acting on the gas molecules. The phase-space distribution function $f$ depends on $(\mathbf{r},\mathbf{u})$, the position and velocity of a gas molecule. Collisions between the gas molecules are modeled by the hard-sphere collision integral $J_{HS}$, see [7]. In [8] it was shown that the problem considered is insensitive to the specific choice of the collision integral, with the variable hard-sphere term giving very similar results.

The Boltzmann equation is solved using a Monte-Carlo scheme, a variant of the DSMC method [9, 10]. For that purpose, the unit cell denoted by the rectangle in Fig. 1 is considered. Periodic boundary conditions are assigned to the boundaries between two unit cells. At all solid surfaces the diffuse reflection boundary condition with complete thermal accommodation is used [7]. The lower surface has a temperature of $T_0 = 723.15\,\mathrm{K}$. The temperature of the upper surface corresponds to the sublimation temperature of dry ice at a pressure of 1 bar, i.e. $T_{subl} = 194.55\,\text{K}$. The key parameter determining the flow regime is the Knudsen number Kn, being the ratio of the mean-free path of the gas molecules and a characteristic length scale. The latter is chosen as $B + H/2$, the former is of the order of 65 nm for carbon dioxide at a mean temperature of 450 K and a pressure of 1 bar. In order to vary Kn, different initial densities are assigned in the simulations. While the model geometry is essentially two-dimensional, all three velocity components of the molecules are computed. A uniform Cartesian grid with quadratic cells is used for tracking the molecules. For $\text{Kn} \geq 1$ a fixed cell size is chosen, whereas for $\text{Kn} < 1$ the cell size is proportional to the molecular mean-free path. Simulations are carried out initializing between 20 and 50 molecules per computational cell using a constant time step. To obtain the velocity field and the shear stress at the upper wall (the boundary of the Leidenfrost solid), averaging over at least $10^5$ time steps is performed. As far as the geometry of the computational domain is concerned, a situation characteristic for experimental studies of Leidenfrost solids is considered. Specifically, $L = 1500\,\mu\text{m}$, $H = 200\,\mu\text{m}$ and $B = 40\,\mu\text{m}$. The latter corresponds to a disc of dry ice with a diameter of 10 mm and a thickness of about 7 mm.

In Fig. 2 the velocity vectors of the computed flow field in a region around the tips of the ratchet are shown for two different Knudsen numbers, $\text{Kn} = 0.06$ and $\text{Kn} = 1$. Since the velocity scales are quite different for these two Knudsen numbers, the vectors were rescaled appropriately for better visualization. There are two counterrotating vortices, a large one forming along the vertical wall and a small one along the incline. When the Knudsen number decreases, the vortices shrink and move closer to the tip of the fin. By contrast, for larger Knudsen numbers the vortices expand and fill an increasing portion of the space between the two surfaces. The CPU time requirements for simulations at small Knudsen numbers are considerable, since very small (compared to the molecular velocity scale) velocities have to be extracted by averaging over a large ensemble of particles and a large number of time steps. Therefore, the smallest Knudsen number that could be considered with the available computational resources while still keeping the statistical fluctuations at a tolerable level was 0.015.

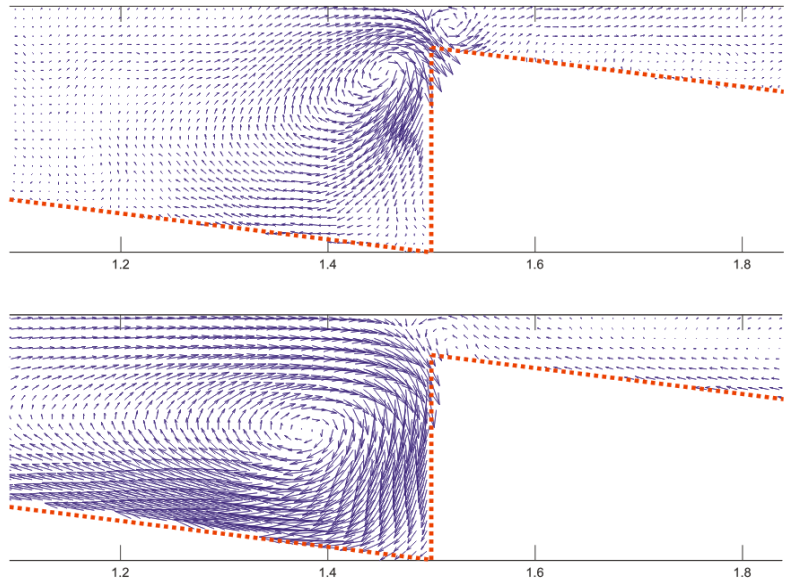

Fig. 2. Velocity vectors in a region around the tips of the ratchet for $Kn = 0.06$ (top) and $Kn = 1$ (bottom). For better visualization, the vectors are scaled differently in both cases.

Apparently, the computed flow field is significantly more complex than the simple thermal creep flow described by Würger [6]. Würger, referring to a seminal paper by Maxwell [11], suggested that along the surface of the ratchet a thermal creep flow establishes, having a velocity of

$$u_c = \frac{3}{4} \nu \frac{\nabla_\parallel T}{T}, \qquad (2)$$

where $\nu$ is the kinematic viscosity of the gas, $T$ its temperature and the gradient is to be taken along the wall. He concluded that along the inclined facets of the ratchet, a thermally-driven flow down the incline develops, with a main component pointing in positive $x$-direction (c.f.

Fig. 1). This flow should drag the Leidenfrost solid along with it, leading to the experimentally observed direction of propulsion. The flow field displayed in Fig. 2 does not correspond to such a simple picture of thermal creep flow. Despite the vortices that cannot be derived from the boundary condition given in Eq. (2), it is apparent that the flow along the incline is in a direction opposite to what is expected from the arguments presented in [6].

The discrepancies between the results presented here and those obtained in [6] can be explained as follows. First of all, effective boundary conditions such as that of Eq. (2) are only applicable to surfaces whose radius of curvature is large compared the thickness of the Knudsen layer, being of the order of the molecular mean free path. At sharp corners phenomena appear that are no longer captured by such effective boundary conditions. Instead, flow patterns termed edge flows develop that often dictate the flow field [12, 13]. The two vortices close to the tips of the ratchet in Fig. 2 nicely illustrate the dominance of thermal edge flow. The second reason for the discrepancies between the present results and the prediction of Würger lies in the fact that in [6] the thermal creep boundary condition was applied in an erroneous manner. The temperature gradient was evaluated one mean free path away from the solid surface. However, the Maxwell boundary condition requires the temperature gradient to be evaluated right at the surface, resulting in a vanishing thermal creep along an isothermal surface [14]. Instead, an isothermal surface can induce a so-called thermal-stress slip flow if the isotherms of the temperature field inside the gas are not parallel to the surface [13]. The direction of the thermal-stress slip flow in the situation studied here is exactly opposite to the thermal creep flow. This is especially reflected in the flow field obtained for $Kn = 1$, where a significant thermal-stress slip flow develops along the incline, augmenting the vortex at the left and weakening the vortex at the right.

To decide which effect this quite complex flow pattern has with respect to propelling the Leidenfrost solid, the shear stress at the upper surface needs to be evaluated. Perhaps surprisingly, even if the flow along the inclines of the ratchet goes into the "wrong" direction, it is redirected through the vortex structure and produces a net shear stress resulting in a total shear force at the upper surface pointing in positive *x*-direction. In that context the vortex at the left dominates over the vortex at the right. Besides the augmentation/weakening of the vortices by the thermal-stress slip flow, the orientation of the isotherms with respect to the two surfaces meeting at a tip of the ratchet decides about the strength of the thermal edge flow. At the vertical parts of the surface the isotherms inside the gas deviate more strongly from the wall orientation than along the inclines, therefore a stronger edge flow develops along the vertical sections.

Fig. 3 shows the nondimensionalized average shear stress $\bar{\tau}_{zx}$ at the upper surface as obtained from the Monte-Carlo simulations. Nondimensionalization was done via dividing by the average pressure evaluated at the upper surface. The shear stress shows the characteristic behavior of transport processes predominantly occurring in the transition flow regime, with a maximum around $\text{Kn} = 1$. For the chosen geometry at a pressure of 1 bar we have $\text{Kn} \approx 4.7 \cdot 10^{-4}$, a value too small to consider in the framework of the Monte-Carlo method. However, taking into account that $\lim_{\text{Kn} \to 0} \bar{\tau}_{zx} = 0$ [14], the average shear stress at very small Knudsen numbers can be determined approximately by interpolation. For that purpose, the data points for the seven smallest Knudsen numbers shown in Fig. 3 were used to compute an interpolation curve going through (0,0), as displayed in the inset of the figure. From this curve a scaling of the shear stress $\bar{\tau}_{zx} \propto \text{Kn}^{1.61}$ is obtained. The scaling law allows computing an approximate value of the thermally induced shear stress corresponding to realistic situations studied in experiments.

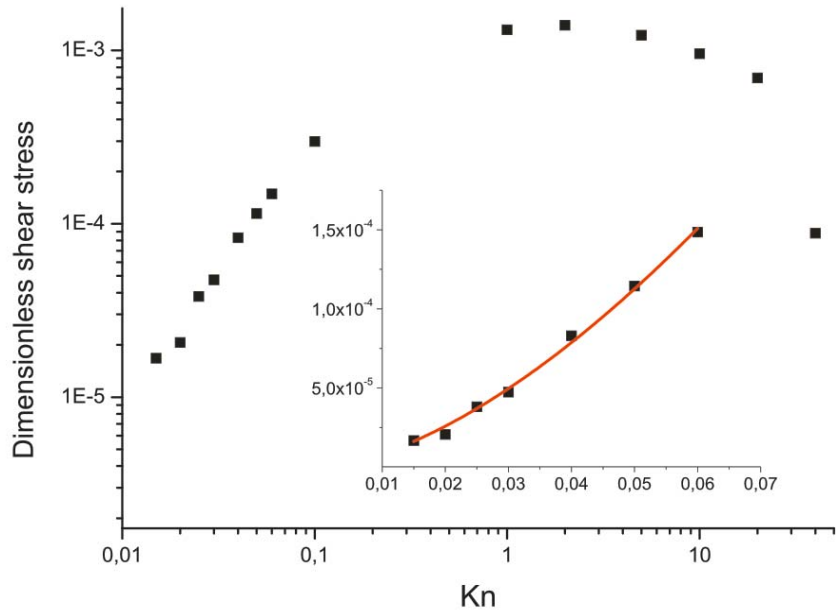

FIG. 3. Dimensionless average shear stress acting on the Leidenfrost solid as a function of Knudsen number. The inset shows a blow up of the data points at small Knudsen numbers together with a power-law fit to the data.

At that point it needs to be mentioned that thermally-driven flows represent only one aspect of the fluid dynamic phenomena contributing to the propulsion of Leidenfrost solids on ratchet. In a separate letter [15] it is shown that the pressure-driven flow due to the production of vapor from the solid gives rise to a net force that compares favorably with experimental results. Typical Reynolds numbers of the pressure-driven flow lie between 1 and 10. Reynolds numbers within that range are small enough to use the superposition principle –strictly valid only for Stokes flow [16]– in an approximate manner. In other words, to a reasonable approximation the two flow phenomena, pressure-driven and thermally-driven flow, may be considered independently, provided that the Reynolds number of the thermally-driven flow does not exceed that of the Stokes flow. To judge the significance of thermally-driven transport, the corresponding shear force onto the solid may be compared to that derived from pressure-driven flow. For a given ratchet geometry, the relative importance of thermally-driven transport depends on the value of the parameter *B*. A rather thick platelet of dry ice (diameter 10 mm, thickness ≈ 7 mm), as considered before, corresponds to $B = 40\,\mu m$. If the thickness is reduced to about 1.7 mm, *B* increases to 100 μm. For the former case, the thermally-induced shear stress amounts to about 2.1 % of the stress due to the pressure-driven flow. For the latter case the relative importance of the thermally-induced stress is larger, amounting to about 13.6 % of the value due to pressure gradients. It may seem counterintuitive that the relative importance of the effects originating from the thermally-driven flow increases when *B* increases, since the corresponding stresses scale as $\bar{\tau}_{zx} \propto \mathrm{Kn}^{1.61}$. However, owing to the combined effects of the decreasing evaporation rate and the decreasing shear stress for fixed evaporation rate, the stresses building up due to the pressure-driven evaporative flux decrease even faster.

One of the results of the analysis of thermally-driven flows is that, owing to the isothermal boundary condition at the surface, no thermal creep flow is induced. Isothermal walls correspond to an infinite thermal conductivity of the substrate, so the question arises what will happen if realistic material properties are considered. Corresponding experiments are usually performed on metal surfaces. Owing to the finite thermal conductivity $\lambda$ of the substrate, a temperature gradient will form along the surface of the ratchet, with the tips being colder than the troughs. In order to estimate the magnitude of the temperature gradient along the incline, the conjugate heat transfer problem was solved using the finite-element software COMSOL Multiphysics. The heat conduction equation was solved inside the solid and inside the gas, being carbon dioxide. The upper surface, the boundary of the Leidenfrost solid, was assumed to be isothermal, while a second parallel isothermal surface at a distance of $2H + B$ below the upper surface served as the lower boundary of the computational domain. The same geometry parameters ($L = 1500\,\mu m$, $H = 200\,\mu m$ and $B = 40\,\mu m$) as before were considered. Even in the presence of evaporation-driven flow, typical Péclét numbers are of the order of 1.

Hence, since the goal is to provide an estimate of the corresponding creep flow, it is justified to neglect convective heat transfer. The problem was solved both for a high thermal conductivity (copper, $\lambda \approx 370\,\text{W}/(\text{m}\cdot\text{K})$) and a low thermal conductivity metal (stainless steel, $\lambda \approx 30\,\text{W}/(\text{m}\cdot\text{K})$). From the resulting temperature gradients along the inclines of the ratchet, the thermal creep velocity is obtained by means of Eq. (2). The latter can be translated into a dimensionless shear stress acting on the Leidenfrost solid using $H$ as a length scale. For the Knudsen number corresponding to 1 bar ($\text{Kn} \approx 4.7\cdot 10^{-4}$) the resulting shear stress values are about $1.2\cdot 10^{-11}$ (copper) and $1.5\cdot 10^{-10}$ (steel). These are much smaller than the average dimensionless stress resulting from the thermally-driven flows shown in Fig. 2, amounting to about $6.4\cdot 10^{-8}$. Therefore, it follows that even when taking into account the finite thermal conductivity of the ratchet substrate, the resulting thermal creep flow can be neglected compared to the phenomena discussed above, especially the thermal edge flow developing at the tips of the ratchet.

In conclusion, it can be stated that thermally-driven flows are not completely insignificant for the propulsion of Leidenfrost solids on ratchets, but they only play a minor role when compared to the pressure-driven flow due to the sublimation of the solid [15]. It should be noted that situations are conceivable (beyond the scope of Leidenfrost phenomena) in which the upper wall in Fig. 1 is simply a solid wall with no mass flux emerging from it. Such situations have been considered in [8]. Especially at large Knudsen numbers specific choices of the wall boundary conditions for the reflection of molecules result in substantial momentum and mass fluxes between the two surfaces. Those may be utilized in energy-conversion applications or for the construction of novel micropumps. In that sense, while the problem studied in this article is only of limited relevance for Leidenfrost objects, it may be important in a different context.